\documentclass{aa}  
\usepackage{natbib}
\bibpunct{(}{)}{;}{a}{}{,} 

\usepackage{graphicx}
\usepackage{txfonts}
\usepackage{hyperref}

\newcommand{\teff}  {T$_\mathrm{eff}$}

\usepackage{amstext}

\begin{document}

\title{Is the activity level of HD\,80606 influenced by its eccentric planet?}

\author{P. Figueira\inst{1},
        A.~Santerne\inst{1},
        A.~Su\'{a}rez Mascare\~{n}o\inst{2},
        J.~Gomes~da~Silva\inst{1},
        L.~Abe\inst{3},
        V.~Zh.~Adibekyan\inst{1},
        P.~Bendjoya\inst{3},
        A.~C.~M.~Correia\inst{4,5},
        E.~Delgado-Mena\inst{1},
        J.~P.~Faria\inst{1,6},
        G.~Hebrard\inst{7, 8},
        C.~Lovis\inst{9},
        M.~Oshagh\inst{10},
        J.-P. Rivet\inst{3},
        N.~C.~Santos\inst{1,6},
        O.~Suarez\inst{3},
        A.~A.~Vidotto\inst{11}
        }

   \institute{Instituto de Astrof\' isica e Ci\^encias do Espa\c{c}o, Universidade do Porto, CAUP, Rua das Estrelas, PT4150-762 Porto, Portugal\\
     \email{pedro.figueira@astro.up.pt}   
     \and
    Instituto de Astrof\'{i}sica de Canarias, E-38205 La Laguna, Tenerife, Spain
    \and
     Laboratoire J.-L. Lagrange, Université de Nice Sophia-Antipolis, CNRS, Observatoire de la Cote d'Azur, F-06304 Nice, France 
    \and
    CIDMA, Departamento de F\'isica, Universidade de Aveiro, Campus de Santiago, 3810-193 Aveiro, Portugal
    \and
    ASD, IMCCE-CNRS UMR8028, Observatoire de Paris, 77 Av. Denfert-Rochereau, 75014 Paris, France
    \and
    Departamento de F\'{i}sica e Astronomia, Faculdade de Ci\^{e}ncias, Universidade do Porto, Portugal
    \and
    Institut d'Astrophysique de Paris, UMR7095 CNRS, Universit\'e Pierre \& Marie Curie, 98bis boulevard Arago, 75014 Paris, France 
     \and
    Observatoire de Haute-Provence, CNRS, Universit\'e d'Aix-Marseille, 04870 Saint-Michel-l'Observatoire, France
    \and
    Observatoire Astronomique de l'Universit\'{e} de Gen\`{e}ve, 51 Ch. des Maillettes, - Sauverny - CH1290, Versoix, Suisse
    \and 
    Institut f\"ur Astrophysik, Georg-August-Universit\"at,  Friedrich-Hund-Platz 1, 37077 G\"ottingen, Germany 
    \and 
    School of Physics, Trinity College Dublin, The University of Dublin, Dublin-2, Ireland
    }

   \date{}

 
  \abstract
   {}
   {Several studies suggest that the activity level of a planet-host star can be influenced by the presence of a close-by orbiting planet. Moreover, the interaction mechanisms that have been proposed, magnetic interaction and tidal interaction, exhibit a very different dependence on orbital separation between the star and the planet. A detection of activity enhancement and characterization of its dependence on planetary orbital distance can, in principle, allow us to characterize the physical mechanism behind the activity enhancement.}
   {We used the HARPS-N spectrograph to measure the stellar activity level of HD\,80606 during the planetary periastron passage and compared the activity measured to that close to apastron. Being characterized by an eccentricity of 0.93 and an orbital period of 111\,days, the system's extreme variation in orbital separation makes it a perfect target to test our hypothesis.}
   {We find no evidence for a variation in the activity level of the star as a function of planetary orbital distance, as measured by all activity indicators employed log($R'_{HK}$), H$_\alpha$, NaI, and HeI. None of the models employed, whether magnetic interaction or tidal interaction, provides a good description of the data. The photometry revealed no variation either, but it was strongly affected by poor weather conditions.} 
   {We find no evidence for star-planet interaction in HD\,80606 at the moment of the periastron passage of its very eccentric planet. The straightforward explanation for the non-detection is the absence of interaction as a result of a low magnetic field strength on either the planet or the star and of the low level of tidal interaction between the two. However, we cannot exclude two scenarios: \textit{i)} the interaction can be instantaneous and of magnetic origin, being concentrated on the substellar point and its surrounding area, and \textit{ii)} the interaction can lead to a delayed activity enhancement. In either scenario, a star-planet interaction would not be detectable with the dataset described in this paper.}

  \authorrunning{P. Figueira et al.}

  \maketitle
  \titlerunning{Is the activity level of HD\,80606 influenced by its eccentric planet?}
%

\section{Introduction}\label{Sec:Intro}

Since the first attempts at detecting extrasolar planets, dedicated surveys have avoided active stars. The photometric and spectroscopic variability of these stars introduce both stochastic and periodic variations in the time series. These photometric and radial velocity (RV) star-induced signals can reach a level similar to that created by an extrasolar planet, and when persistent, can even mimic a planetary signal \cite[e.g.,][]{2007A&A...474..293B, 2008A&A...489L...9H, 2010A&A...513L...8F, 2014A&A...566A..35S}. Yet, it has for long been argued that the activity level of a star might be enhanced by the presence of a planet around it; if that is indeed the case,  we might be heavily biasing our scrutiny of the planetary population by neglecting active stars.

One of the first works to propose that stars hosting extrasolar planets were more active than non-planet hosts was \cite{2008ApJ...687.1339K}. The authors studied the X-ray activity of stars hosting close-in giant planets and found that these were more active than stars with planets at a wider separation. The idea initially gathered some support, but was later refuted by more detailed statistical analysis, showing that the correlation was in all likelihood created by selection effects \citep{2010A&A...515A..98P, 2014A&A...565L...1P}. \cite{2015ApJ...799..163M} recently studied the possible correlation between activity and the most common proxies for iteration derived from orbital parameters, namely $M_p/a^2$ or $1/a$, and found no evidence for a correlation. Given the difficulty in characterizing the different biases at work, several studies chose to focus on individual planet-hosting stars and searched for a correlation between the host activity level and the orbital phase of a close-in massive planet. The first work to report such a link was that of \cite{2005ApJ...622.1075S}, but subsequent observations failed to recover the signal \citep{2008ApJ...676..628S, 2013A&A...552A...7S}. 
Approaching the problem from a different angle, \cite{2014A&A...567A.128P} showed that the X-ray luminosity of WASP-18 is more than two orders of magnitude lower than expected for a star of its age and mass. The authors proposed that the the orbiting planet is responsible for this discrepancy and that a close-in planet can significantly reduce the activity of the host star.

The unavoidable conclusion was that the activity enhancement, if present, was of an {\it \textup{\textup{on/of}f}} nature and very difficult to separate from the intrinsic stellar variation. The literature on possible correlations between planetary orbital phase and stellar activity is abundant but inconclusive \citep[e.g.,][]{2011A&A...525A..14L, 2012A&A...543L...4L, 2011ApJ...741L..18P, 2015ApJ...805...52P}. On the other hand, theoretical and numerical models \citep{2009A&A...505..339L, 2011ApJ...733...67C, 2015A&A...578A...6M, 2015ApJ...815..111S} have long suggested that magnetic star-planet interaction intermediated by reconnection events could release energetic particles that would travel toward the star, triggering potentially measurable activity enhancement events \citep[but see, e.g., ][]{2012A&A...544A..23L}. Very recently, \cite{2016ApJ...820...89F} found tentative evidence for star-planet interaction from the variation of several FUV lines (NV, CIV, SiIV). The same authors found a correlation between the magnitude of the interaction effect and the line formation temperature of high temperature (upper chromosphere or corona) lines. In spite of the abundant literature in the subject, the picture remained unclear.

Very interestingly, and adding another layer of complexity to the subject at hand, it was shown that the activity of a planet-host star as measured using the log($R'_{HK}$) indicator correlates inversely with the planetary surface gravity \cite[for a full description of the correlation and its robustness we refer to   ][]{2010ApJ...717L.138H, 2014A&A...572A..51F, 2016OLEB..tmp...24F, 2015ApJ...812L..35F}. \cite{2014A&A...572L...6L} explained this correlation by invoking selective absorption at the core of CaII H+K line by obscuring material released by the evaporation of these giant planets, a scenario that was previously proposed by \cite{2012ApJ...760...79H} for the extreme case of WASP-12. 

Clearly, an unambiguous planet-induced stellar activity enhancement would provide an important benchmark. As such, the work of \cite{2015ApJ...811L...2M} was received with enthusiasm. It reported that the stellar activity of the close-by HD\,17156 seems to show a dependence on the planetary orbital phase of its eccentric planet. A significant enhancement in activity, measured using the log(R'$_{HK}$) index, was detected close to periastron passage; very close in time, an increase in X-ray luminosity was apparent. This enhancement is well in line with the idea that star-planet interaction is expected to have a strong dependence on the star-planet separation \citep{2000ApJ...533L.151C}, and as such be maximum near periastron. However, the evidence was not as strong as predicted: of several optical spectra obtained close to the periastron passage, only one showed a clear enhancement in activity.

To understand the nature of star-planet interaction and test the hypothesis of proximity-enhanced activity, we observed the exoplanet HD\,80606\,b as it crossed the periastron of its orbit. With an eccentricity of 0.93 and a semi-major axis of 0.449\,AU, HD\,80606\,b reaches a distance $d$ from the star of only 
0.03\,AU, a value to be compared with that of HD\,17156\,b, which passes at a distance of 0.05\,AU from its host star at periastron ($e$\,=\,0.68; $a$\,=\,0.162). Owing to the strong dependence of magnetic interaction on orbital distance, which scales with $d^{-2}$ \citep{2000ApJ...533L.151C}, we can expect an activity enhancement 2.8 times stronger on HD\,80606 than on HD\,17156. On the other hand, if we consider a tidal interaction effect, which is expected to be proportional to the bulge size \citep[e.g.,][]{2014A&A...570L...5C}, we have
\begin{equation}
 h_{tide} \propto \frac{M_{p}}{M_{s}}\frac{R_s^4}{d^3}\,,
\end{equation}

in which $M_p$ and $M_S$ are the masses of the planet and of the star, respectively, $R_S$ is the mean radius of the star, and $d$ is the distance between the centers of the star and the planet. If we insert in this equation the system data of HD\,80606 ($M_p$\,=\,3.94\,$M_{Jup}$, $P$\,=\,111.4\,days, $M_s$\,=\,0.98\,$M_{\odot}$, $R_s$\,=\,0.98$R_{\odot}$) and HD\,17156  ($M_p$\,=\,3.19\,$M_{Jup}$, $P$\,=\,21.2\,days, $M_s$\,=\,1.28\,$M_{\odot}$, $R_s$\,=\,1.51$R_{\odot}$), we conclude that the signal induced by HD\,80606\,b is $\sim$33\% stronger than the one triggered by HD\,17156\,b. In this comparison, we assume that the underlying properties of the stars, such as the mass of the convective envelope or the stellar magnetic field properties, are the same; these assumptions may not be justified, but the calculus shows us that an effect is expected to be detectable on HD\,80606.


In summary, HD\,80606 and its eccentric planet present a very favorable scenario to repeat the experiment of \cite{2015ApJ...811L...2M} in HD\,17156, and in more favorable conditions \citep[see also ][]{2010AJ....140.1929L, 2011MNRAS.414.1573V}. In principle, when observations obtained at (or close to) periastron and apastron
are considered, the steep dependence on star-planet separation allows us to probe the physical mechanism behind the enhancement: magnetic, tidal, or other. In the rather extreme case of a magnetically induced enhancement that is present throughout the whole phase, the activity ratio between periastron and apastron will be of $\sim$\,760, as dictated by the $d^{-2}$ dependence; for a tidal-induced enhancement, it will be of a factor of $\sim$400, following the $d^{-3}$ dependence. The difference in the measurable output can enable us to understand, or at least narrow down the options about the mechanism at work without having to compare the effect on different hosts. 

In Sect.\,2 we describe the observations acquired, and in Sect.\,3 we interpret and discuss our results. In Sect.\,4 we conclude on our work, answering our initial questions. \\


\section{Observations and results}

\subsection{Spectroscopy with HARPS-N}

We used the HARPS-N spectrograph \citep{2012SPIE.8446E..1VC}, mounted at the TNG telescope, to acquire high-resolution optical spectra of the star HD\,80606 as its planetary companion passed close to apastron and periastron. The spectra were used to measure the RV variation and the activity level of the star. The dataset was acquired through a TNG DDT program (program ID: A32DDT4) and consisted of two points obtained close to apastron and four points close to periastron. The requested set of observations was composed of one more point close to the periastron and one more point close to the apastron, but poor weather conditions made their acquisition impossible. The S/N ratio of the spectra ranged from 14 to 25 when measured at 390\,nm, close to the CaII H\&K lines, and from 70 to 125 when measured at the center of the strongest orders. Using the ephemerides of \cite{2010A&A...516A..95H}, we were able to determine the periastron time with an error of approximately 13 min. This precision and the favorable periastron passage time allowed us to observe HD\,80606\,b much closer to periastron than HD\,17156\,b; 13 min corresponds to $\Delta \phi\,<\,8.0 \times 10^{-4}$ for HD\,80606\,b, while \cite{2015ApJ...811L...2M} reported an enhancement at $\Delta \phi\,\approx\,1.0 \times 10^{-2}$ for HD\,17156\,b. It is important to note that for HD\,80606\,b there will be no periastron passage visible from TNG telescope that will occur in night time or close to it in the next five years, making this dataset a very important one.

\begin{figure*}
\centering
\includegraphics[width=15cm]{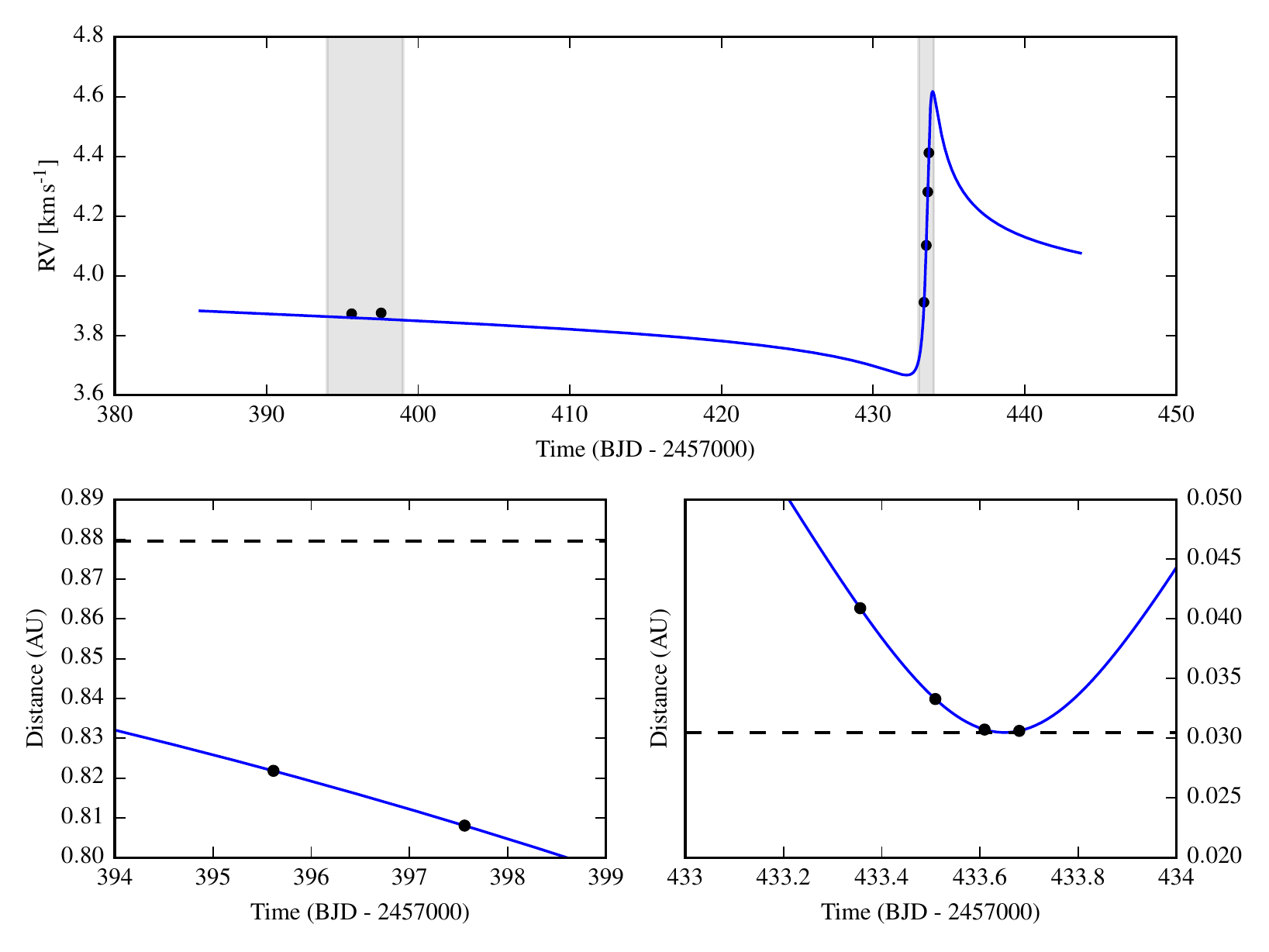}
\caption{RV and orbital distance of HD\,80606\,b for the six data points acquired, overplotted on the orbit derived from the parameters and ephemerides of \cite{2010A&A...516A..95H}. The upper plot shows the variation in RV as a function of baricentric Julian date, and the lower plots represent the orbital distance of the companion when close to apastron ({\it left}) and periastron ({\it right}). The dashed lines represent the apastron and periastron distance, for visual reference.}
\label{Fig:RV}     
\end{figure*}

The spectra were processed using the HARPS-N pipeline, very similar to that of its predecessor and twin HARPS \citep{2003Msngr.114...20M}. The RV were calculated by cross-correlation with a weighted binary mask \citep{1996A&AS..119..373B, 2002A&A...388..632P}, a procedure that is now standard. We derived the indicators $BIS$, $BIS^+$, $\Delta V$, and $V_{span}$ along with the cross-correlation function (CCF) $FWHM$ \citep[for a description of each one of the these
we refer to][ and references therein]{2013A&A...557A..93F}\footnote{The program used to calculate the indicators was introduced in \cite{2014A&A...566A..35S} and is available from \url{https://bitbucket.org/pedrofigueira/line-profile-indicators}.} to check for activity-induced RV signals. The RV error estimated by the pipeline and line profile indicator values are presented in Table\,\ref{Tab:RV}, in which we also list the computed orbital distance. In Fig.\,\ref{Fig:RV} we plot the RV variation as a function of time overplotted on the orbit predicted using the ephemerids and orbital parameters of \cite{2010A&A...516A..95H}. We note that RV and associated indicators are calculated for the baricenter of the solar system. The time of observation is also calculated for the same point in space and is termed baricentric Julian date. The difference between baricentric Julian date and Julian date is given by the light travel time, modulus relativistic effects, and as such is of $\approx$8 minutes. Since this time precision is never attained in the comparison between RV and photometric series,  we treat baricentric Julian date and Julian date as the same in the remainder of the paper.

\begin{table*}
\centering
\caption{RV, associated uncertainty, orbital distance, and line-profile indicators for the HARPS-N dataset.}
\label{Tab:RV}
\begin{tabular}{c|cc|c|ccccc}
\hline\noalign{\smallskip}
jdb [day] & RV\,[km/s] & $\sigma_{\mathrm{RV}}$\,[km/s]& a\tablefootmark{\dagger}\,[AU] & $BIS$\,[km/s] & $BIS^+$\,[km/s] & $\Delta V$\,[km/s] & $V_{span}$\,[km/s] & $FWHM$\,[km/s] \\
\noalign{\smallskip}\hline\noalign{\smallskip}
2457395.615 &   3.87263 & 0.00220       & 0.8218 & -0.03284     & -0.04956        & 0.04983       & -0.02705      & 7.12858       \\
2457397.561 &   3.87535 & 0.00311       & 0.8080 & -0.03355     & -0.05164        & 0.05043       & -0.02653      & 7.13108       \\
2457433.356 &   3.91107 & 0.00306       & 0.0408 & -0.03507     & -0.05225        & 0.05501       & -0.03027      & 7.13636       \\
2457433.509 &   4.10185 & 0.00221       & 0.0333 & -0.03431     & -0.05474        & 0.05193       & -0.02869      & 7.13942       \\      
2457433.609 &   4.28092 & 0.00217       & 0.0307 & -0.03144     & -0.04757        & 0.04975       & -0.02759      & 7.13823       \\
2457433.680 &   4.41195 & 0.00240       & 0.0306 & -0.03166     & -0.04543        & 0.04933       & -0.02754      & 7.13644       \\
\noalign{\smallskip}\hline
\end{tabular}
\tablefoot{\tablefoottext{\dagger}{Calculated using the ephemerides of \cite{2010A&A...516A..95H}.}}
\centering
\end{table*}

The RV residuals of these RV data points relative to the published orbit were 21\,m/s, a value well in excess of the estimated RV uncertainties. This prompted us to rederive the parameters for this orbit. We analyzed the HARPS-N data presented here along with the RV data of \cite{2010A&A...516A..95H} and references therein (from the HiReS, HRS, and SOPHIE spectrographs). The analysis was performed assuming a Keplerian orbit and using the Markov chain Monte Carlo algorithm of the \texttt{PASTIS} software, fully described in \cite{2014MNRAS.441..983D}. We used uninformative priors for the orbital eccentricity, argument of periastron, systemic velocity, and radial velocity amplitude. We used a normal distribution for the orbital period matching the results reported by \cite{2010A&A...516A..95H} but widening the width by 100 to avoid biasing the results by too much. The Spitzer photometry obtained by \cite{2010A&A...516A..95H} was not included in the fit, but we used their reported value for the transit epoch and duration as prior for the analysis.
The refined orbital parameters of HD80606\,b and their 63.8\% confidence intervals are presented in Table\,\ref{Tab:parameters}, so that they can be considered in future works. When used for the orbital fitting, these updated parameters exhibit RV residuals at a level of 4\,m/s, fully compatible with the average uncertainty on the RV points. 

With an orbital eccentricity of 0.93, HD\,80606\,b is the second most eccentric planet known, following HD\,20782\,b. The availability of high-precision RV measurements spanning many years motivated us to attempt to fit the precession of the periastron, $\dot \omega$. Several effects contribute to this variation, such as general relativity, polar oblateness, and tidal deformation \citep[for a review see][]{2011CeMDA.111..105C}. Of these, general relativity
dominates with an estimated value $\dot \omega \approx 0.0006^\circ$/yr. In our analysis we obtain $\dot \omega = 0.027 \pm 0.031\,^\circ$/yr, which means that in spite of the large data set and its precision,we are not yet able to detect this effect.


\begin{table}
\centering
\caption{Orbital parameters rederived with {\tt PASTIS}.}
\label{Tab:parameters}       
\begin{tabular}{cc}
\hline\noalign{\smallskip}
K & 473.0 $\pm$ 2.3\,m/s \\
P & 111.43734 $\pm$ 1.7$\times10^{-4}$\,d \\
$T_{tran}$ & 55210.64203 $\pm$ 9.9$\times10^{-4}$ \\
a/Rs & 97.3 $\pm$ 1.5 \\
$R_p/R_s$ & 0.10009 $\pm$ 6.1$\times10^{-4}$ \\
$i$ & 89.267 $\pm$ 0.017$^\circ$ \\
$e$ & 0.93166 $\pm$ 6.1 \\
$\omega$ & 301.21 $\pm$ 0.17$^\circ$ \\
$\dot \omega$ & 0.027 $\pm$ 0.031\,$^\circ$/yr \\
$\lambda$ & 52$^\circ$ $^{+25}_{-14}$ \\
$\gamma$& 3.7881 $\pm$ 2.0$\times10^{-3}$\,km/s \\
$v.\sin{i}$ & 1.31 $\pm$ 0.35\,km/s \\
\hline
ELODIE RV jitter & 11.6 $\pm$ 2.1\,m/s \\
HARPS-N RV jitter & 5.5 + 5.4 - 2.4\,m/s \\
HARPS-N RV offset & -145.7 $\pm$ 5.2\,m/s \\
HET RV jitter & 2.0$^{+2.2}_{-1.4}$\,m/s \\
HET RV offset & 3.8074 $\pm$ 2.8$\times10^{-3}$\,km/s \\
Keck1 RV jitter & 5.81 $\pm$ 0.97\,m/s \\
Keck1 RV offset & 3.9733 $\pm$ 2.3$\times10^{-3}$\,km/s \\
Keck2 RV jitter & 1.87 $\pm$ 0.48\,m/s \\
Keck2 RV offset & 3.9709 $\pm$ 2.1$\times^{-3}$\,km/s \\
SOPHIE1 RV jitter & 1.35 $\pm$ 1.2\,m/s \\
SOPHIE1 RV offset & -126.0 $\pm$ 2.3\,m/s \\
SOPHIE2 RV jitter & 3.18 $\pm$ 0.85\,m/s \\
SOPHIE2 RV offset & -113.7 $\pm$ 2.2 m/s \\
\noalign{\smallskip}\hline
\end{tabular}
\centering
\end{table}

The activity level was measured in each spectrum using four independent activity indicators. First we calculated the log($R'_{HK}$) index following the description of \cite{1984ApJ...279..763N} and the implementation of \cite{2011arXiv1107.5325L}. This was obtained by post-processing the HARPS-N spectra using the YABI interface\footnote{Accessible from \url{http://ia2-harps.oats.inaf.it:8000/login/?next=/}.} The flux in the H$_\alpha$ line was calculated using a rectangular band shape of 0.16\,nm around 656.2808\,nm, and two continuum band shapes of 1.075\,nm around 655.087\,nm and 0.875\,nm around 658.031\,nm, following the procedure described in \cite{2011A&A...534A..30G} and \cite{2015MNRAS.452.2745S}. The index is then given by the flux in the first bandpass normalized by the sum of the fluxes in the other two. The errors were calculated following standard error propagation.

For NaI and HeI we followed the procedure of \cite{2011A&A...534A..30G}, who in turn followed the procedure of \cite[][see their Sect. 5.1 for the index computation, and Sect. 3.1 for the measurement of the continuum]{2007MNRAS.378.1007D} for NaI and \cite{2009A&A...495..959B} (See their Sect. 5.3 for the index computation) for HeI.

\begin{figure*}
\centering
\includegraphics[width=15cm]{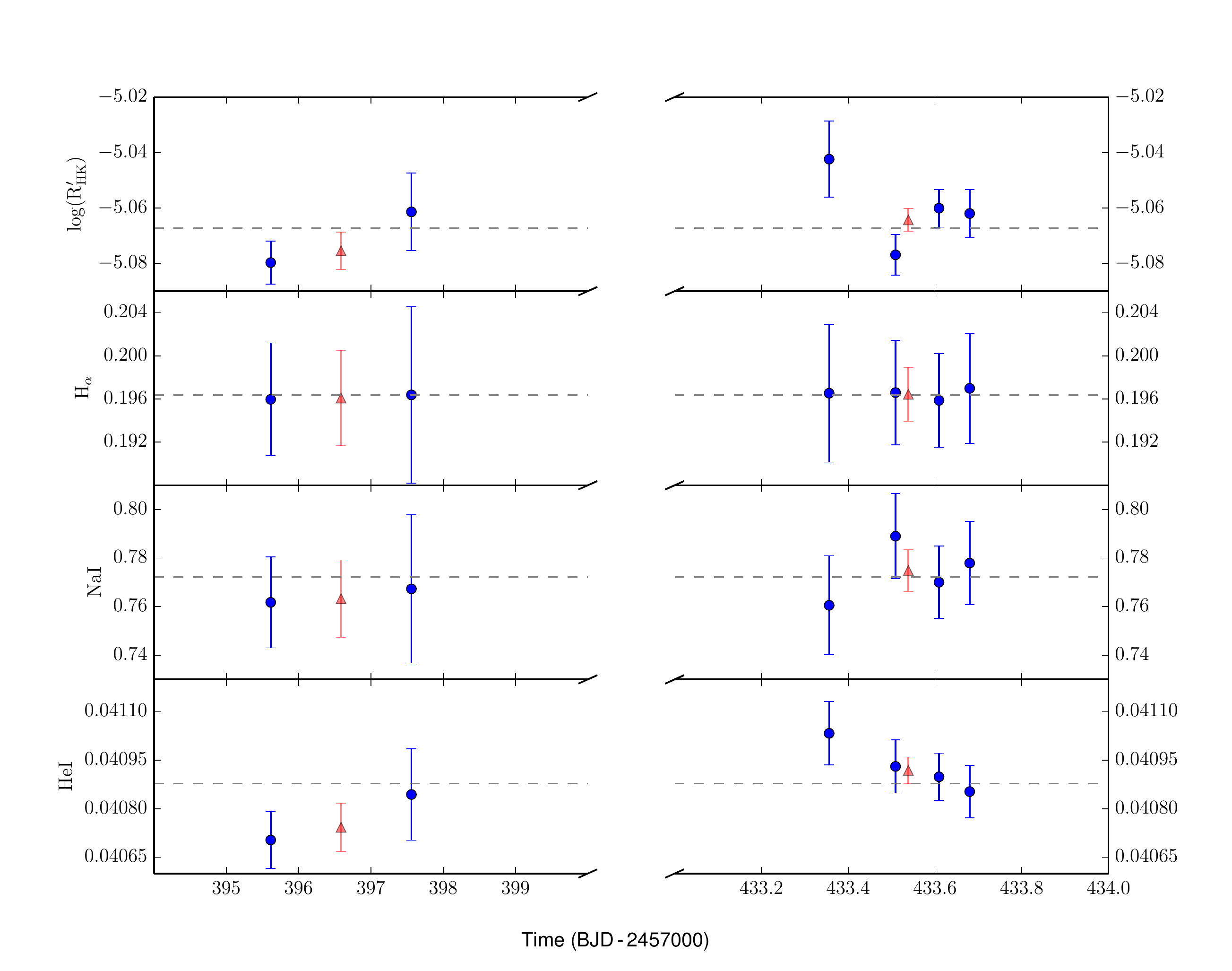}
\caption{Value of the four activity indicators considered: log($R'_{HK}$), H$_{\alpha}$, NaI, and HeI, from top to bottom, as a function of Julian date. The points close to apastron are represented in the left panels and those close to periastron in the right panels. The blue points represent the individual measurements and uncertainties as listed in Table\,\ref{Tab:actindex}. In red we plot the weighted average and associated uncertainty for the points close to apastron and points close to periastron. The dashed line is the weighted average of the six measurements and is a visual guideline for the variation inside the dataset. }
\label{Fig:actindex}     
\end{figure*}

An increase in stellar activity is identified by an increase in the coefficient value for log($R'_{HK}$), H$_\alpha$, and NaI, and a decrease for HeI. We list the value of each indicator and associated uncertainties in Table\,\ref{Tab:actindex} and plot the results in Fig.\,\ref{Fig:actindex}.

\begin{table*}
\centering
\caption{Spectral activity indicators measured for each spectra of the HARPS-N dataset.}
\label{Tab:actindex}       
\begin{tabular}{c|cccc}
\hline\noalign{\smallskip}
jdb [day] &  log$(\mathrm{R}'_{HK})$ & H$_{\alpha}$ & NaI & HeI \\
\noalign{\smallskip}\hline\noalign{\smallskip}
2457395.615   &   -5.080\,$\pm$\,0.008   &   0.762\,$\pm$\,0.019   &   0.196\,$\pm$\,0.005    &   4.070e-02\,$\pm$\,8.808e-05 \\
2457397.561   &   -5.061\,$\pm$\,0.014   &   0.767\,$\pm$\,0.030   &   0.196\,$\pm$\,0.008    &   4.084e-02\,$\pm$\,1.414e-04 \\
2457433.356   &   -5.042\,$\pm$\,0.014   &   0.761\,$\pm$\,0.020   &   0.197\,$\pm$\,0.006    &   4.103e-02\,$\pm$\,9.796e-05 \\
2457433.509   &   -5.077\,$\pm$\,0.007   &   0.789\,$\pm$\,0.018   &   0.197\,$\pm$\,0.004    &   4.093e-02\,$\pm$\,8.210e-05 \\
2457433.609   &   -5.060\,$\pm$\,0.007   &   0.770\,$\pm$\,0.015   &   0.196\,$\pm$\,0.004    &   4.090e-02\,$\pm$\,7.272e-05 \\
2457433.680   &   -5.062\,$\pm$\,0.009   &   0.778\,$\pm$\,0.017   &   0.197\,$\pm$\,0.005    &   4.085e-02\,$\pm$\,8.118e-05 \\
\noalign{\smallskip}\hline
\end{tabular}
\centering
\end{table*}

The availability of high-resolution and spectra with high signal-to-noise ratio (S/N) prompted us to refine the analysis of the stellar parameters published in \cite{2004A&A...415.1153S} that is based on UES spectra. The current analysis followed a similar procedure, but making use of the larger FeI and FeII line-list of \cite{2008A&A...487..373S}. The current analysis yields \teff\,=\,5617$\pm$52\,K, log(g)\,=\,4.48$\pm$0.07, and [Fe/H]\,=\,0.33$\pm$0.03, which is very similar to the published values, even if more precise. We also estimated the abundances of the most common species by applying a differential line-by-line analysis relative to a high S/N reference spectra of the Sun taken with HARPS, as detailed in \cite{2016A&A...591A..34A}. The abundances and associated uncertainties are presented in Table\,\ref{Tab:abundances}. The lithium content is very low, as expected for a star with such a \teff \,\citep[e.g.,][]{2014A&A...562A..92D} and the abundances in general reveal a star that follows the Galactic chemical evolution trend. Recently, \cite{2015A&A...579A..52N} showed that the [Y/Mg] ratio can be used to estimate stellar ages for solar-like stars. This result was later confirmed by \cite{2016arXiv160405733T}, who provided an empirical relation between the two parameters. The [Y/Mg] -- age relation from \cite{2016arXiv160405733T} suggests an age of 5.7$\pm$2.0\,Gyr for our target.

\begin{table}
\centering
\caption{Abundances of the most common species.}
\label{Tab:abundances}       
\begin{tabular}{c|c}
\hline\noalign{\smallskip}
[X/H] & A\,$\pm\,\sigma$ \\
\noalign{\smallskip}\hline\noalign{\smallskip}
CI & 0.310 $\pm$        0.030   \\      
OI &    0.481 $\pm$     0.093   \\      
NaI     &       0.453 $\pm$     0.044 \\
MgI     &       0.356 $\pm$     0.036 \\
AlI     &       0.349 $\pm$     0.033 \\
SiI     &       0.340 $\pm$     0.019 \\
SI &    0.340 $\pm$     0.060   \\      
CaI     &       0.221 $\pm$     0.041 \\
$<$Sc$>$        &       0.429 $\pm$     0.028 \\
$<$Ti$>$        &       0.355 $\pm$     0.031 \\
VI      &       0.505 $\pm$     0.062 \\
$<$Cr$>$        &       0.337 $\pm$     0.034 \\
MnI     &       0.450 $\pm$     0.054 \\
CoI     &       0.434 $\pm$     0.025 \\
NiI     &       0.385 $\pm$     0.022 \\

CuI &   0.520 $\pm$     0.067   \\
ZnI &   0.341 $\pm$     0.043   \\
SrI &   0.269 $\pm$     0.035   \\
YII &    0.307 $\pm$    0.068   \\
ZrII &   0.365 $\pm$    0.074   \\
BaII &   0.163 $\pm$    0.035   \\
CeII &   0.384 $\pm$    0.085   \\
NdII &   0.335 $\pm$    0.050   \\

\hline
A(Li) & $<$0.80 \\
\noalign{\smallskip}\hline
\end{tabular}
\centering
\end{table}

\subsection{Photometric monitoring}

We performed photometric observations of HD\,80606 using the one-meter Omicron telescope of the C2PU observing facility at the Observatoire de la C{\^o}te d'Azur \footnote{UAI code\,:~$010$, lat$=43.7537^\circ$~N, lon$=6.9230^\circ$~E, alt=1270\,m.} from mid-January to mid-February 2016. 

This instrument has a $F/3.2$ prime focus with a three-lense coma corrector (Wynne corrector). The sensor used is a SBIG STX16803 camera ($4096\times4096$ pixels, $9\times9\,\mu$m each) with a pixel scale of $0.56$ arcsec/pixel. All the observations were made with a photometric B~filter to maximize the contrast between the photosphere and the energetic flare \citep{2014Ap.....57...77M} that could be induced by the planet. Each individual frame corresponds to $3$ to $6$~seconds exposure time, depending on the sky transparency. Frames with pixels above $42\,000$ ADU were rejected to avoid photometric nonlinearity effects.

\begin{figure*}
\centering
\includegraphics[width=12cm]{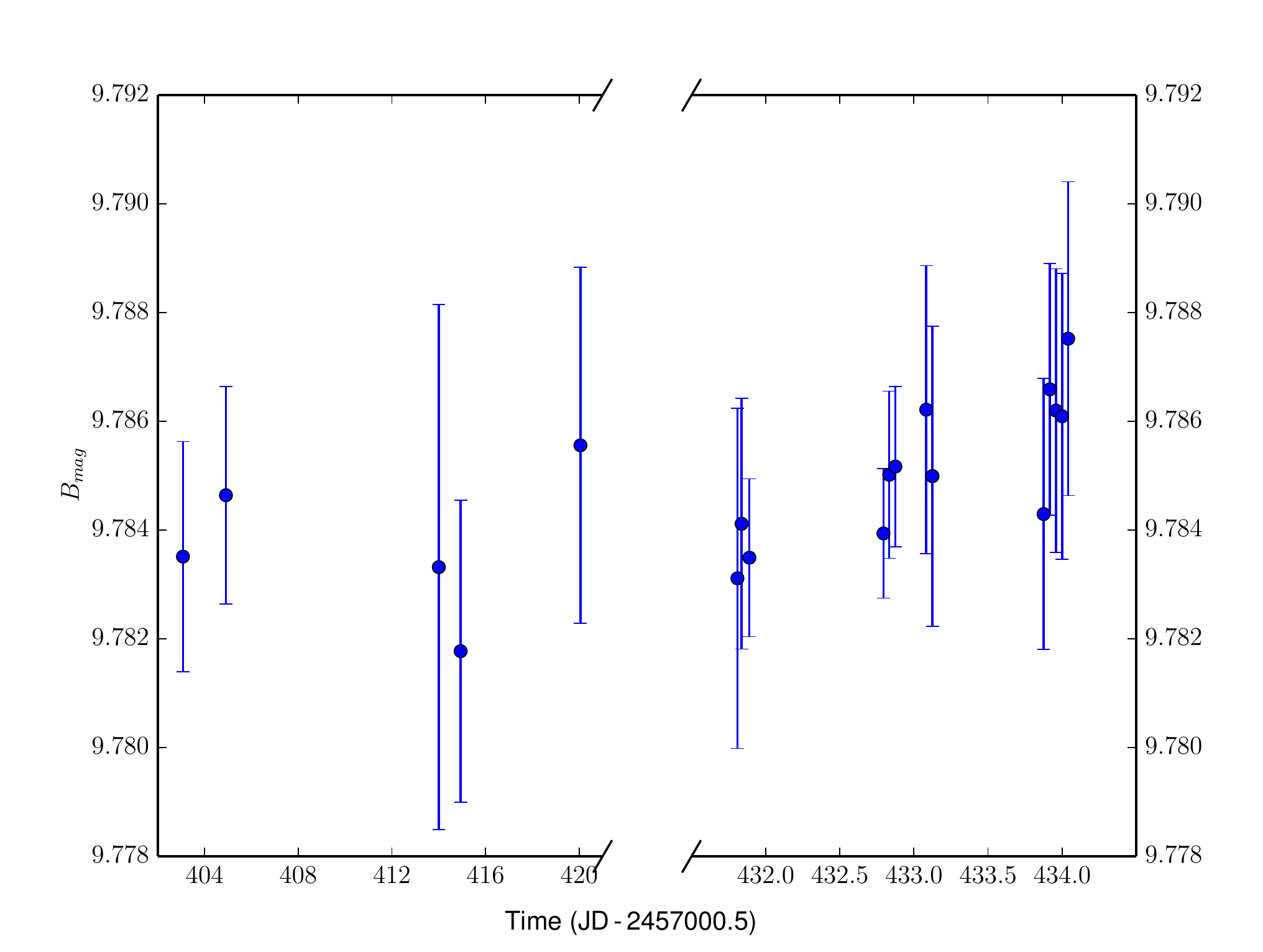}
\caption{B~magnitude versus modified UTC Julian date ($JD-2400000.5$) for the complete photometric observation campaign of HD80606 on the C2PU one-meter telescope (2016~Jan.~16 to Feb.~15). The left panels represents the variation before periastron and the right panel the variation during periastron night and adjacent nights.}
\label{Fig:photometry}     
\end{figure*}

The data were processed with {\sl AstroImageJ} vers.~$3.2.0$, a Java-based image processing software. First, the science images were calibrated following standard procedures (median dark-frame subtraction, median unit-normalized flat-field division, outlier removal). Then, relative aperture photometry was applied, using the neighboring star HD\,80607 as a reference star. The aperture photometry was performed with inner and outer radii for the sky annulus of $19$ and $25$ pixels, respectively. The radius of the aperture disk was allowed to fluctuate according to the FWHM of the stellar image in each individual frame. Consequently, the B magnitudes provided in Table\,\ref{Tab:photometry} and Fig.~\ref{Fig:photometry} are computed assuming that the B magnitude of HD\,80607 is 9.937.
For the nights of February 13-15, the telescope was dedicated to the observations of HD80606. A set of five preliminary observations was performed from January 16 to February 1, in which a series of integrations of HD\,80606 were inserted between other observing programs. Unfortunately, these nights were affected by poor weather conditions, namely strong atmospheric turbulence and unstable transparency. Consequently, the photometric accuracy is poor,  $2-3$ millimag, with an average of 2.5 millimag. During these preliminary runs the star did not show any noticeable variability.

During the nights before, during, and after the predicted periastron of HD\,80606b, the C2PU Omicron telescope was fully dedicated to the observation of the host star. Bad seeing conditions and several cloudy periods interrupted the observations during those three nights, and consequently the data set is not continuous. Table\,\ref{Tab:photometry} and Fig.~\ref{Fig:photometry} list and present the B~magnitude as a function of the UTC modified Julian date (MJD) for the whole campaign; the observations close to periastron, on February~13-15, are displayed in the right panel. For the night of February 13, the data are sparse because
of highly unstable weather conditions. Each data point on the graph corresponds to the average over a continuous observing period ($37.3$~min, $17.5$~min, and $38.8$~min, respectively). For the nights of February~14 and 15, each data point corresponds to the average over a continuous observing period
of nearly one hour.

Compared to the B~magnitudes measured during the preliminary observation runs, the star did not exhibit any clear brightness variation during the expected periastron epoch at the level of photometric accuracy achieved.

\begin{table}
\centering
\caption{B~magnitude versus UTC Julian date for the complete photometric observation campaign of HD\,80606 on the C2PU one-meter telescope. The magnitude was calculated assuming as constant the magnitude of the reference star HD\,80607 (see text for details).}
\label{Tab:photometry}       
\begin{tabular}{c|c}
\hline\noalign{\smallskip}
JD [day] &  $B_{mag}$ \\
\noalign{\smallskip}\hline\noalign{\smallskip}

2457403.575   &   9.784\,$\pm$\,0.002  \\
2457405.409   &   9.785\,$\pm$\,0.002  \\
2457414.505   &   9.783\,$\pm$\,0.005  \\
2457415.437   &   9.782\,$\pm$\,0.003  \\
2457420.554   &   9.786\,$\pm$\,0.003  \\
2457432.310   &   9.783\,$\pm$\,0.003  \\
2457432.338   &   9.784\,$\pm$\,0.002  \\
2457432.390   &   9.783\,$\pm$\,0.001  \\
2457433.295   &   9.784\,$\pm$\,0.001  \\
2457433.333   &   9.785\,$\pm$\,0.002  \\
2457433.375   &   9.785\,$\pm$\,0.001  \\
2457433.583   &   9.786\,$\pm$\,0.003  \\
2457433.625   &   9.785\,$\pm$\,0.003  \\
2457433.625   &   9.785\,$\pm$\,0.003  \\
2457434.375   &   9.784\,$\pm$\,0.002  \\
2457434.417   &   9.787\,$\pm$\,0.002  \\
2457434.458   &   9.786\,$\pm$\,0.003  \\
2457434.500   &   9.786\,$\pm$\,0.003  \\
2457434.541   &   9.788\,$\pm$\,0.003  \\

\noalign{\smallskip}\hline
\end{tabular}
\centering
\end{table}

\section{Interpretation and discussion}

\subsection{Is the stellar activity enhanced?}

A quick analysis of Table\,\ref{Tab:actindex} or Fig.\,\ref{Fig:actindex} will promptly reveal that no clear activity enhancement was detected using any indicator. For log$(\mathrm{R}'_{HK})$, H$_{\alpha}$, NaI, and HeI the scatter (measured using the standard deviation and considering Bessel's correction to yield an unbiased estimator) is only 1.39, 0.07, 0.54, and 1.16 times the average uncertainty on the activity index, respectively. The variation is then comparable to and compatible with the average measurement uncertainty.

We can use the dependence of the activity enhancement on the
planet-star separation as described in Sect.\,\ref{Sec:Intro} to evaluate whether an interaction mechanism is at work. Different enhancement mechanisms will lead to a different dependence: a purely magnetic interaction will scale with $d^{-2}$, while a purely tidal interaction will scale with $d^{-3}$. We can therefore calculate the linear coefficients (a,\,b) such that $Act_m(d)$\,=\,$a_m \times d^{-2} + b_m$ or $Act_t(d)$\,=\,$a_t \times d^{-3} + b_t$. We performed a weighted least-squares regression\footnote{Using
the {\tt python} library {\tt statsmodel}.} using as weights the inverse of the square of the activity indicator's uncertainty. The values obtained for $a_m$ and $a_t$ for each indicator, along with the associated uncertainties derived from the covariance matrix inversion, are presented in Table\,\ref{Tab:WLS}. For all the different computed indicators, the slope values are always comparable with their own uncertainties; as expected, no clear evidence for star-planet interaction is found. The probability of false alarms from F-statistics is always in the range 20-60\%, showing that the data are unlikely to be described by this dependence.

\begin{table}
\centering
\caption{Slope coefficient and uncertainties from weighted least-squares fitting assuming a purely magnetic or purely tidal variation of activity with orbital phase, as discussed in the text.}
\label{Tab:WLS}       
\begin{tabular}{c|cc}
\hline\noalign{\smallskip}
indicator & $a_m$ & $a_t$ \\
\noalign{\smallskip}\hline\noalign{\smallskip}

log$(\mathrm{R}'_{HK})$ &       9.761e-06 $\pm$ 1.10e-05 & 2.892e-7 $\pm$ 3.40e-7   \\

H$_{\alpha}$    &  2.805e-07 $\pm$ 4.63e-07 & 7.511e-9 $\pm$ 1.43e-8 \\

NaI     & 1.352e-05 $\pm$ 9.63e-06 & 4.091e-7$\pm$ 2.92e-7 \\

HeI     & 1.216e-07 $\pm$ 9.51e-08 & 3.019e-9 $\pm$ 3.08e-9 \\

\noalign{\smallskip}\hline
\end{tabular}
\centering
\end{table}

\begin{figure*}
\centering
\includegraphics[width=12cm]{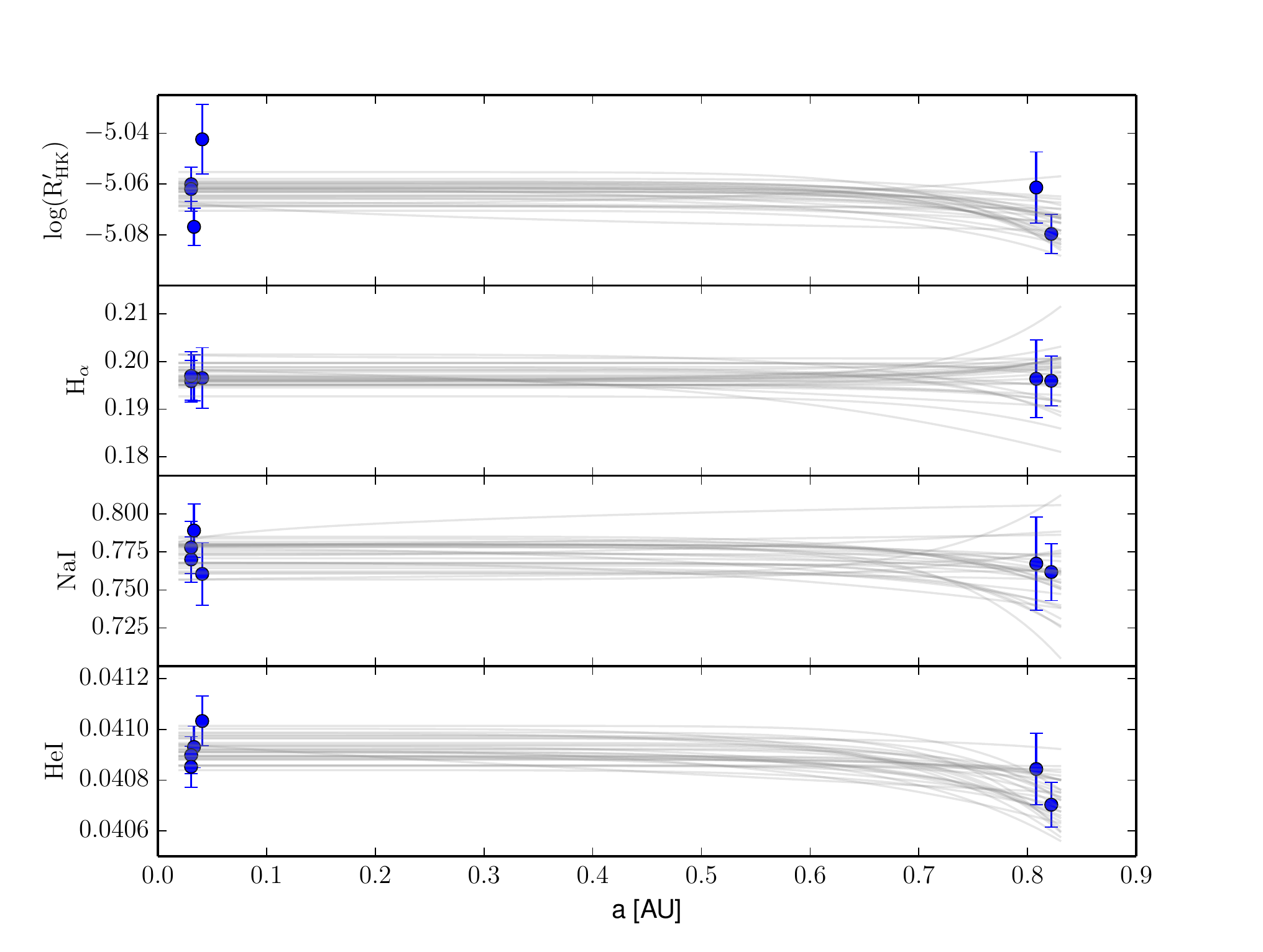}
\caption{Variation of the four indicators as a function of orbital distance, and 30 models drawn randomly from the posterior distribution.  }
\label{Fig:MCMC}     
\end{figure*}

However, for an interaction that is neither purely magnetic nor purely tidal, we expect a dependence on orbital distance on a different power than -2 or -3. It is then informative to fit a function of the type $Act(d) = A \times d^n +c$ in which $A$, $n$, and $c$ are the parameters to fit. We used a Levenberg-Marquardt algorithm to attempt a fit \footnote{To do it we used the computer language {\tt python} and the library {\tt scipy.optimize}; the function used ({\tt curve\_fit}) is a wrapper around $MINPACK$'s $lmdif$ and $lmder$ algorithms.}. We soon realized that the results were highly dependent on the initial guesses for the parameters. In particular, when each of the values [-1, 0, 1] are considered as first guess for \textit{n}, we obtain completely different final values for the parameter. To understand the effect of our prior assumption on each parameter, we used the Bayesian framework to estimate the probability distribution of each parameter. We used the computer language {\tt python} and the library {\tt pyMC} \citep{2015ascl.soft06005F} for Bayesian inference and Monte Carlo posterior sampling. We concluded, once again, that the results depended heavily on the priors, and in particular on the allowed range for $n$. In Fig.\ref{Fig:MCMC} we plot 30 samples drawn randomly from the derived posterior distribution, showing that the data do not constrain these models.

A simple way of testing whether activity is enhanced is to compare the results of the (weighted) average of the two points near apastron and the four points near periastron, as representative of the star under its minimum and maximum activity, respectively. We can then evaluate the probability that the difference is zero, or negative, as opposed to an activity enhancement, which would be materialized in a positive difference. The distribution of a difference of two normally distributed variables $X$ and $Y$ with means and variances ($\mu_X$, $\sigma_X^2$) and ($\mu_Y$, $\sigma_Y^2$) is a Gaussian itself
\begin{equation}
P_{X-Y}(u) = \frac{e^{-[u - (\mu_X - \mu_Y)]^2/(2(\sigma_x^2 + \sigma_y^2)}}{\sqrt{2 \pi (\sigma_x^2 + \sigma_y^2)}}\,,
\end{equation}

with $\mu_{X-Y} = \mu_X - \mu_Y$ and $\sigma^2_{X-Y} = \sigma_X^2 + \sigma_Y^2$. We can then calculate the probability that the difference between the periastron and apastron is $P_{X-Y} \le 0$. This procedure was employed for log$(\mathrm{R}'_{HK})$, H$_{\alpha}$, NaI; for HeI, and since an increasing activity leads to a decrease in the activity index value, we considered the difference between apastron and periastron. The values of $P_{X-Y} \le 0$ for the four indicators are 7.9\%, 47.2\%, 26.1\%, and 98.0\%. This means that the probability of a non-enhancement is always too high to be discarded.

With a log($R'_{HK}$) below -5.0 dex, HD\,80606 is a remarkably inactive star. For such a low level of activity, an increase in activity leads to an increase in absorption of H$_{\alpha}$. This is expected to create an anticorrelation between the activity as measured in CaII or NaI and that of H$_{\alpha}$, as demonstrated for FGK activity evolution over long timescales by \cite{2014A&A...566A..66G}. Such a correlation is not supported by the data, again reinforcing the absence of a detectable activity variation. On the other hand, a visual inspection of Fig.\,\ref{Fig:actindex} reveals what appears to be a monotonic decrease in activity for the periastron data of HeI. This variation also seems to correlate with a similar variation for the $\Delta V$ indicator. However, we have to be very careful with such interpretations. We performed an independent analysis of four activity indicators and five line-profile indicators, each with six points. The presence of a monotonic variation in a subset of our data or correlation between two pairs of variables can be the result of multiple hypothesis testing. We therefore
did not consider the monotonic decrease of HeI during periastron, or its correlation with $\Delta V$, as conclusive evidence for enhanced activity. Moreover, the higher average value of HeI during periastron shows that if a planetary-induced HeI absorption is present at periastron, the scatter and activity at apastron has to be due to other effects, such as the activity level of the star, which undermines our argument. 

\subsection{Understanding the non-detection}

None of the datasets we gathered shows evidence for a stellar activity enhancement as the planet crosses its periastron. While the photometry was undoubtedly compromised by poor weather conditions, the spectroscopy data were of high-quality: the instrument is known for its high fidelity and the reported uncertainties of 0.007-0.014\,dex in log($R'_{HK}$) attest to the quality of the derived indicator as a telling example. Concerning photometric precision, the average error bar was $\approx$2.5\,mmag, and the peak-to-peak measurement $\approx$6\,mmag (as measured in the normalized light curve). We used the {\tt SOAP2.0} software \citep{2014ApJ...796..132D}, to check which type of activity structure, that is, spots or plages, could reproduce the RV scatter of 4\,m/s while generating the peak-to-peak photometric variation of 6\,mmag. There is a natural degeneracy between the temperature of an active region and its filling factor; several combinations of the two parameters can induce the same RV and photometric variation. To estimate whether the filling factor for each structure type is comparable to that of stars of the same type, we fixed the temperature of the active regions to the values measured for solar-type stars. We used for plages a (positive) temperature difference of 250\,K \citep{2010A&A...519A..66M} and for spots a (negative) temperature difference of 1250\,K \citep{2005LRSP....2....8B}. We concluded that to reproduce the RV scatter and photometric peak-to-peak variation we can only use a stellar spot; a plage inducing a photometric variation of this amplitude will generate an RV variation of about \,40 m/s. We are able to reproduce these observations with a single spot with a filling factor of 0.45\% and a temperature contrast of 1250 K. The maximum filling factor for Sun-spots in the low-activity phase is of 1\%, and its temperature constrast can reach up to 2500\,K \citep{2003A&ARv..11..153S}. This shows that the filling factor of the postulated spot is well within the typical range for a quiet star like our own Sun, showing that, once again, the dataset is consistent with no appreciable variation of activity.

The impressive rise in RV shows that the data acquisition was well timed, and an ephemerids error therefore cannot explain the observed data. There are several possible explanations for the non-detection.

The simplest explanation is that in spite of the close proximity of the planet at periastron, the planet does not trigger any activity enhancement in the star. Alternatively, if enhancement does occur, it is possible that this was below our detection sensitivity for all the indicators. In particular, for the case of magnetic interactions, we need both planet and star to have a sufficiently strong magnetic field for the magnetic reconnection event to dissipate a detectable amount of energy.

The fraction of the stellar hemisphere facing the planet and observable by us from our vantage point is given by $f = [1+\cos(E)]/2$, with $E$ being the angle between the line of sight and the star to its host planet. The angle $E$ is defined by $\cos(E) = \sin(\omega+\nu)\times\sin(i)$ , in which $\omega$ is the periastron angle, $\nu$ the true anomaly, and $i$ the inclination of the system. From the definition of $\omega$ and $\nu$, when the planet is at the periastron we have that $\nu$\,=\,0. Using the data from Table\,\ref{Tab:parameters}, we obtain f\,=\,7\%. This means that we are observing only a very small fraction of the closest hemisphere to the star, the one that is, in principle, more susceptible to an activity enhancement from magnetic interaction. Clearly, a magnetic activity enhancement that would be instantaneous and localized close to the substellar point would not be detectable from our vantage point. This would not be the case for a tidal effect, which would manifest itself on opposite sides of the star, and thus be accessible from our vantage point.

Another possible explanation for the lack of activity enhancement is that there is a phase shift between the substellar phase and the phase at which activity enhancement occurs. This has been suggested before for other systems \citep{2008ApJ...676..628S} and might occur as a result of the complex configuration of stellar magnetic fields \citep{2006MNRAS.367L...1M, 2010MNRAS.406..409F}. It is not unusual for stellar magnetic field lines to become twisted \citep{2012MNRAS.423.3285V}. In this case, the line connecting the star and the planet would not be anchored at the substellar point, but would instead be shifted to a different stellar longitude and/or latitude. 
Finally, we note that transfers of energy between the star and the planet only occur when the planet is orbiting within the stellar Alfv\'enic surface \citep{2015ApJ...815..111S}. Therefore, understanding the realistic distribution of stellar wind particles and magnetic fields are of utmost importance to model and quantify star-planet interactions \citep{2014MNRAS.438.1162V}. Clearly, characterizing the topology of the magnetic fields present in HD\,80606, using Doppler imaging or spectropolarimetric observations, could provide an extra piece to the puzzle posed by these observations.

Both HD\,80606 and HD\,17156 have a low activity level, with previously reported values for log$(\mathrm{R}'_{HK})$ of -5.061 and -5.022, respectively \citep{2014A&A...572A..51F}. We can associate a low activity level with a low intrinsic stellar magnetic field and consequently a low likelihood for interaction events, but this does not explain the absence of a detection on HD\,80606 when one was reported on HD\,17156. It is also well-known that stellar activity cycles can affect the measured activity level of FGK stars \citep[e.g.,][]{2014A&A...566A..66G}. However, as a result of the long periods and low amplitude of such effects, the effect on the study performed here is expected to be very low. Finally, two planets lie above the correlation locus between log$(\mathrm{R}'_{HK})$ and planetary surface gravity; they are located in the upper left quadrant of Fig.\,1 of \cite{2014A&A...572A..51F} because of their high surface gravity and low activity level. Therefore there is no evidence for the influence of an evaporation or absorption effect on the measured activity. While the two planets have an higher eccentricity than average, there is no clustering, or preferential disposition of the eccentric planets relative to the locus of the correlation. This means that eccentricity does not seem to work as a hidden variable or to have any effect on the planetary surface gravity {\it \textup{vs.}} stellar activity correlation.


\section{Conclusion}\label{sec:Conclusions}

We used the HARPS-N spectrograph to acquire spectra of the planet-host HD\,80606 as its eccentric planet crossed the periastron of its orbit. The periastron observations were compared to observations taken close to apastron, which would reveal an activity enhancement due to the distance-dependent interaction, as predicted by several theoretical works. In spite of the high fidelity and high S/N of the data, no significant activity variation was identified using four well-established activity indicators: log($R'_{HK}$), H$_\alpha$, NaI, and HeI. The straightforward explanation for the non-detection is the absence of interaction, which in itself can be due to a low magnetic field strength on either the planet or the star. However, we cannot exclude two scenarios: {\it i)} the interaction can be instantaneous and of magnetic origin, being concentrated on the substellar point and its surrounding area, and {\it ii)} the interaction can lead to a delayed activity enhancement. In either scenario, a star-planet interaction would not be detectable with the dataset presented in this paper. A full characterization of the stellar magnetic field of HD\,80606 can help to confirm or exclude these scenarios.


\begin{acknowledgements}
This work was supported by Funda\c{c}\~ao para a Ci\^encia e a Tecnologia (FCT) (project ref. PTDC/FIS-AST/1526/2014) through national funds and by FEDER through COMPETE2020 (ref. POCI-01-0145-FEDER-016886), as well as through grant UID/FIS/04434/2013 (POCI-01-0145-FEDER-007672). This work results from the collaboration of the COST Action TD 1308.
PF and NCS acknowledge support by Funda\c{c}\~ao para a Ci\^encia e a Tecnologia (FCT) through Investigador FCT contracts of reference IF/01037/2013 and IF/00169/2012, respectively, and POPH/FSE (EC) by FEDER funding through the program ``Programa Operacional de Factores de Competitividade - COMPETE''. PF further acknowledges support from Funda\c{c}\~ao para a Ci\^encia e a Tecnologia (FCT) in the form of an exploratory project of reference IF/01037/2013CP1191/CT0001. 
AS is supported by the EU under a Marie Curie Intra-European Fellowship for Career Development with reference FP7-PEOPLE-2013-IEF, number 627202.
ASM acknowledges financial support from the Spanish project MINECO AYA2014-56359-P.
EDM, VZhA and JPF also acknowledge the support from the FCT in the form of the grants SFRH/BPD/76606/2011, SFRH/BPD/70574/2010, and SFRH/BD/93848/2013, respectively. AC acknowledges support from CIDMA strategic project UID/MAT/04106/2013.
MO acknowledges research funding from the Deutsche Forschungsgemeinschaft (DFG , German Research Foundation) - OS 508/1-1. We thank the referee, Luca Fossati, for his careful revision of the manuscript.
The team is indebted to Antonio Magazzu and Emilio Molinario at TNG for their prompt assistance and friendliness, and to Andrew Vanderburg for performing the periastron observations in service mode. PF would also like to warmly thank the HARPS-N management board for allowing a time swap and make observations during periastron possible. We thank everyone who contributed to developing the open-source {\tt python} language and keeping it free, and in particular the developers of the open-source package {\tt pyMC}.

\end{acknowledgements}

\bibliographystyle{aa} 
\bibliography{Mybibliog} 






\end{document}